# Designing Ultra-Flat Bands in Twisted Bilayer Materials at Large Twist Angles without specific degree

Shengdan Tao[1], Xuanlin Zhang[2], Jiaojiao Zhu[3], Pimo He[1], Shengyuan A. Yang[3], Yunhao Lu*[1,2], Su-Huai Wei[4]

[1] *Zhejiang Province Key Laboratory of Quantum Technology and Device, Department of Physics, Zhejiang University, Hangzhou 310027, China*

[2] *State Key Laboratory of Silicon Materials, School of Materials Science and Engineering, Zhejiang University, Hangzhou 310027, China*

[3] *Research Laboratory for Quantum Materials, Singapore University of Technology and Design, Singapore 487372, Singapore*

[4] *Beijing Computational Science Research Center, Beijing 100193, China*

**ABSTRACT**: Inter-twisted bilayers of two-dimensional (2D) materials can host low-energy flat bands, which offer opportunity to investigate many intriguing physics associated with strong electron correlations. In the existing systems, ultra-flat bands only emerge at very small twist angles less than a few degrees, which poses challenge for experimental study and practical applications. Here, we propose a new design principle to achieve low-energy ultra-flat bands with increased twist angles. The key condition is to have a 2D semiconducting material with large energy difference of band edges controlled by stacking. We show that the interlayer interaction leads to defect-like states under twisting, which forms a flat band in the semiconducting band gap with dispersion strongly suppressed by the large energy barriers in the moiré superlattice even for large twist angles. We explicitly demonstrate our idea in bilayer α-$In_2Se_3$ and bilayer InSe. For bilayer α-$In_2Se_3$, we show that a twist angle ~13.2˚ is sufficient to achieve the band flatness comparable to that of twist bilayer graphene at the magic angle ~1.1˚. In addition, the appearance of ultra-flat bands here is not sensitive to the twist angle as in bilayer graphene, and it can be further controlled by external gate fields. Our finding provides a new route to achieve ultra-flat bands other than reducing the twist angles and paves the way towards engineering such flat bands in a large family of 2D materials.

**INTRODUCTION**

The ratio between interaction and kinetic energies is an indicator of the significance of electron correlation effects. In solids, the kinetic energy is embodied by the bandwidth. Hence, systems with flat bands around the Fermi level offer promising platforms for exploring intriguing physics associated with strong electron correlations. Recently, it was discovered that flat bands can emerge in twisted bilayer two-dimensional (2D) materials and lead to correlated insulating and superconducting phases[1],[2], which attracted tremendous research interest[3]-[9]. So far, in the reported systems, the appearance of ultra-flat bands requires the delicate tuning to a very small twist angles. For example, in twisted bilayer graphene, the bandwidth of 5-10 meV is achieved at the magic angle ~ 1.1°[1],[2]. For other systems like 2D transition metal dichalcogenides[10],[11], black phosphorene[12] and BN[13], the similar bandwidth also requires a tiny twist angle. The very small twist angle poses challenge for fabrication. In addition, it leads to a huge unit cell for moiré superlattice, e.g., the lattice constant is increased 39 times for twisted bilayer graphene at $\theta$ =1.47°[14]. This constrains the possibility of making compact and scalable devices based on such systems. Therefore, for both fundamental studies and future applications, it is highly desired to have systems where ultra-flat bands can be achieved at large twist angles.

In the simple tight-binding picture, the bandwidth can be estimated as $W = 2zt$, where $z$ is the number of neighboring sites and $t$ is the hopping strength. Hence, to minimize the bandwidth, the main task is to find a way to suppress $t$. In twisted 2D bilayers, the states that comprise the flat band are defect-like states spatially located around regions with a particular local layer stacking configuration, which form the superlattice structure[15]. Intuitively, to suppress the hopping strength $t$ between the sites of this superlattice, we have two approaches. The first approach is to increase the separation distance between the sites, i.e., to increase the lattice constant of the moiré superlattice. This is in fact the underlying origin that the existing systems have small twist angles, because the period of moiré superstructure increases with decreasing twist angle[16],[17]. The second approach to suppress the hopping is to increase the energy barrier between the sites. In a moiré superlattice, the region around a site and the region between sites correspond to different stacking configurations[18],[19]. Thus, the high energy barrier essentially requires that the defect-like state is more localized and has a large energy difference between the two stacking configurations. If the energy barrier is sufficiently high, we can expect the formation of ultra-flat band even for large twist angles with non-specific degree. This is the idea that we are going to explore in this work.

The general consideration above leads us to a new design principle to achieve ultra-flat bands in twisted 2D bilayers. Our proposal is the following. First, because we want the flat band to be formed within a band gap as a defect-like band, we need a 2D material whose monolayer and bilayer are semiconductors. Second, we request that in the monolayer, the states around the band edge, either the conduction band minimum (CBM) or the valence band maximum (VBM), distributed mainly on the outmost atomic layer and have wave functions extended outward with anisotropic character. For example, in Figure 1a, we illustrate a scenario when the VBM of some monolayer semiconductor is dominated by $p_z$ orbitals, which meets our condition. Now, we construct a bilayer system by orienting the two monolayers such that their band edge states are extended towards the interlayer region (see Figure 1a). Upon twisting, we have a moiré supercell containing regions with different local stacking configurations. Clearly, the stacking configuration strongly affects the overlap between the band edge states from the two layers. For example, the stackings I and II illustrated in Figure 1a will result in a large difference in the coupling between the VBM states. Obviously, stacking II has a stronger coupling, and it leads to a larger energy splitting for the VBM states in the bilayer compared to stacking I (see Figure 1a). Assume that in the twisted bilayer, we have regions with local stacking II forming the superlattice sites, and the regions of stacking I located between the sites, as shown in Figure 1b. Then, the split state indicated by the red line in Figure 1a will form defect-like states trapped at stacking II regions, like orbitals for a large artificial atom. And the hopping between neighboring sites will need to overcome the energy barrier, whose order of magnitude is indicated by the energy difference $\Delta E$ between stackings I and II, as indicated in Figure 1a. Evidently, our prescriptions target to maximize the energy barrier, such that the defect states can be strongly localized and result in a flat band around Fermi level isolated from other bands.

We explicitly demonstrate our idea by first-principles calculations of concrete material examples. Let us first consider 2D α-In$_2$Se$_3$. We are going to see that with our approach, an ultra-flat band with bandwidth comparable to twisted bilayer graphene can be achieved in this system at a large twist angle ~ 13.2˚. And for twist angle at ~

10°, the bandwidth can be further suppressed to ~ 0.9 meV, suggesting even stronger correlation effects than graphene. In addition, for α-In$_2$Se$_3$, the emergence of flat band is also related to the interlayer polarization alignment, and the flat bands can be switched on/off by external electric field.

**RESULTS AND DISCUSSION**

The structure of a monolayer α-In$_2$Se$_3$ is shown in Figure 2a, which consists of five atomic layers in the sequence of Se-In-Se-In-Se and has *P*3*m*1 polar space group (No. 156)[20]. The two indium atomic layers in the structure are AB stacked and have different coordination structures. In one indium layer (the lower In layer in Figure 2a), each In atom is surrounded by six Se atoms, forming InSe$_6$ octahedra. For the other indium layer (the upper In layer in Figure 2a), four Se atoms are present around each In atom, forming InSe$_4$ tetrahedra. The different coordination for the two In layers breaks the inversion symmetry and produces an out-of-plane electric polarization, which has been established in experiment[21],[22]. Figure 2b shows the calculated band structure of monolayer α-In$_2$Se$_3$. It is a semiconductor with almost direct band gap around the Γ point. Focusing on the low-energy bands, they are mostly from the In *s* orbitals and Se *p* orbitals, whereas the main contribution from the In *d* orbitals is away from the Fermi level. In Figure 2b, we plot the projection weight for different atomic orbitals. One can observe that the CBM is mainly from the *s* orbitals of In atoms in the tetrahedra, i.e., the upper In layer in Figure 2a, whereas the states around VBM are mainly from the *p* orbitals of Se atoms in the octahedra. In Figure 2c, one observes that the VBM state at Γ has major contribution from the lowest Se layer. Moreover, the highest valence band state at K point, which is quite close to the VBM energy, is dominated by the $p_z$ orbitals of the lowest Se layer in the octahedra. We can see that these valence band states fulfill our condition: they are located at the outermost atomic layer and extended outward with anisotropic character.

Now, we construct a bilayer of α-In$_2$Se$_3$[23]. According to our proposal, in the bilayer, we should have the octahedra in each layer face to each other (end-to-end polarization alignment), as shown in Figure 3a. Let's first examine two different stacking configurations of the bilayer in the absence of twisting. Focusing on the two Se layers (one from each α-In$_2$Se$_3$ layer) closest to the interlayer region, analogous to those in Figure 1a, we consider the AA stacking where the Se atoms of the two layers are aligned on top of each other and the AB stacking where the atoms from the upper layer is lying on the voids of the lower layer, as illustrated in Figure 3a. Evidently, we choose these two, because they are expected to exhibit the most contrasted interlayer coupling. For the AB stacking, since the two Se layers are horizontally shifted to form a staggered pattern, one can expect that the interlayer coupling for the valence band states is weak. In comparison, the interlayer coupling for the AA stacking should be much stronger. Particularly, this should be the case for the state at K, because the $p_z$ orbitals from the two Se layers have lobes directed exactly towards each other, resulting in a strong interaction. These points are confirmed by the first-principles results in Figure 3c which show the band structures of bilayers with the two stackings. Indeed, the most notable difference is observed at the K point. For AA stacking, there is a large energy splitting at K, which is capable to push the upper valence state to become the new VBM, whereas for AB stacking, the splitting is small and the two states have a degeneracy at K as they constitute the 2D irreducible representation of the $D_3$ group at K. In Figure 3d, we plot the spatial distribution of the states K1 and K2 indicated in Figure 3c. One observes that the strong coupling produces covalent-like quasi-bonding[24] between layers that separates the antibonding state (K1) ~1.2 eV from the associated bonding state (K2). Referenced to the vacuum level, we find that the VBM for the AA stacking has an energy $\Delta E \sim 0.2$ eV higher than that for the AB stacking, which is a significant value and implies a large energy barrier (or deep potential well) for hole carriers when a moiré superlattice with these two stackings are formed.

Next, we add twist to the bilayer. As mentioned, upon twisting, the bilayer will develop a moiré superstructure, and in a moiré supercell, there are regions correspond to different local stacking configurations. For example, in Figure 3b, we show the bilayer with a twist angle $\theta = 13.2°$. The regions with local AA and AB stackings are marked with red and black circles, respectively. According to our discussion, the VBM states for this setup will be mainly distributed in the red circled regions with AA stacking. This is confirmed by our calculation result in Figure 4a. In Figure 4, we also show the results for several other twist angles, ranging from $\theta = 32.2°$ to $9.43°$ along the twist-axis I shown in Figure 3a (Very similar results of other twist-axes are shown in the Supporting Information). Indeed, due to the large energy barrier that suppresses the hopping, localized states around AA stacking regions and the associated flat band already appear at $\theta = 32.2°$. The bandwidth decreases with decreasing twist angle: $W=$

46.7 meV, 32.7 meV, 9.4 meV and 0.9 meV at $\theta$=32.2°, 21.8°, 13.2° and 9.43°, respectively. One observes that the flatness in twisted bilayer graphene (~ 5-10 meV) with $\theta \sim 1.1°$ [1],[2] can be readily achieved in the current system at a large twist angle $\theta$ =13.2°. Remarkably, at the twist angle of 9.43°, the bandwidth shrinks to be less than 1 meV, which suggests strong correlation effects and surpasses all reported systems so far.

In our proposed design principle, we request that the band edge states should have large distribution at the outermost atomic layer and extend towards the interlayer gap with anisotropic character. This condition is satisfied by the valence band states in α-$In_2Se_3$ but not the conduction band states. Hence, we do not expect flat bands from the conduction band states at relatively large twist angles. This is also explicitly confirmed by our calculation. Since the CBM states in monolayer α-$In_2Se_3$ are mainly from the In atoms in the tetrahedra, for the bilayer configurations in Figure 3a where the tetrahedra are away from the interlayer region, the interlayer interaction has little effect on the CBM (see Figure 3c). In Figure 5a, we change the layer orientation of α-$In_2Se_3$ bilayer such that the tetrahedra of the two layers are facing to each other (head-to-head polarization alignment) and compare the two types of stackings AA and AB. One finds that the conduction bands have little dependence on the stacking orders, as shown in Figure 5b. The maximum CBM energy difference for the two stackings is ~ 0.03 eV, an order of magnitude smaller than the VBM difference in Figure 3c. This means the energy barrier (or the potential well) is low, which cannot effectively suppress the kinetic energy for electron carriers. Thus, at similar twist angles, there is no flat band resulting from the conduction band states, as shown in Figure 5c.

The essential physics here can be understood from the famous Kronig-Penny model[25] (see Figure 6a). It is well known that in this model, the bandwidth for a particular band follows the relation $W \sim 1/(V_0 ba)^2$, where $V_0$ is the height of energy barriers, $b$ and $a$ are the width of the barrier and the well, respectively. Thus, to suppress the bandwidth, one can (1) increase $a$ and $b$, or equivalently, the lattice period (For twisted bilayers, this is done by decreasing the twist angle, which increase the moiré superlattice period); (2) increase the energy barrier $V_0$, which is the approach adopted in this work. This qualitative behavior is general. In Figure 6b we plot the bandwidth for the topmost valence band versus the maximum VBM energy difference $\Delta E$ between different stackings (which plays the role of $V_0$) for several twisted bilayer 2D semiconductors at a fixed twist angle of 9.43°. One can indeed observe the general trend of decreasing $W$ with increasing $\Delta E$.

As mentioned, the monolayer α-$In_2Se_3$ has an out-of-plane electric polarization and it can be switched by an applied electric field, so it is a ferroelectric. This ferroelectric character is helpful for making the band edge states asymmetric across the atomic layers, however, it is not necessary for our proposal. As an example, consider 2D InSe, which has no electric polarization. Its VBM states again have large distribution extending towards the interlayer region and have the oriented character of $p$ orbitals. This leads to the large maximum VBM difference ~ 0.16 eV between AA and AB stackings, thus resulting in ultra-flat band at relatively large twist angles. From the result in Figure 6c and d, one can see that the bandwidth in twisted InSe bilayer can be suppressed to be less than 10 meV at $\theta = 9.43°$.

We note that the ferroelectric character of α-In$_2$Se$_3$ may offer a method to effectively control the flat band. It has been shown that bilayer α-In$_2$Se$_3$ naturally favors the antiparallel interlayer polarization with the end-to-end configuration[26], such as those in Figure 3. Hence, the twisted bilayer with ultra-flat band can be realized as a ground state configuration in a device setup. Then, by applying a gate field, one can align the polarization of the two layers. In the process, the layer with polarization against the *E* field undergoes a structural transition which effectively flip the layer upside down. Clearly, the transition would suppress the interlayer coupling for the VBM states. As a result, the flat band would disappear, as illustrated by Figure S6 in the Supplementary. Thus, in a dual gated device of twisted bilayer α-In$_2$Se$_3$ (see Figure 5d), we can not only control the doping of the ultra-flat band, but also control its presence or not.

**CONCLUSION**

In conclusion, above results reveal a new mechanism to obtain flat bands other than reducing twist angle. Given a sufficient energy difference between band edges of different stacking orders, the flat band can form at large twist angles without specific degree. As no specific magic angles are required, it is possible to obtain flat bands for some 2D materials at any twist angle. Our results also explain the different critical twist angle for the emergence of flat band in a serials of 2D vdW materials and have important implications for the future search of flat-band characters in moiré structures.

**COMPUTATIONAL DETAILS**

Our first-principles calculations were carried out within the density functional theory based on the projected augmented wave (PAW)[27],[28] pseudopotentials as implemented in the Vienna Ab-initio Simulation Package (VASP)[29]. The generalized-gradient approximation (GGA) in the form of Perdew-Burke-Ernzerhof (PBE)[30] functional was used. The cutoff of the plane wave basis was set to guarantee that the absolute energies are converged to a few meV. The Grimme's method was employed to incorporate the effects of vdW interactions and different vdW corrections give the same conclusion. The vacuum region was set larger than 15 Å to minimize artificial interactions between images. The 2D Brillouin zone was sampled by the Γ-centered Monkhorst-Pack k-point mesh and it is 8×8 for the untwisted α-$In_2Se_3$ bilayer, as well as 2×2 and 1×1 for the moiré superlattices with $\theta$ =32.2°, 21.8° and $\theta$ =13.2°, 9.43° respectively. The configurations of twisted bilayer were constructed according to accident angular commensurations. The positions of atoms were fully relaxed until the Hellmann-Feynman force on each atom was less than 0.02 eV/Å. The convergence criteria for energy was set to $10^{-5}$ eV. Spin-orbital coupling (SOC) has little effect on the results and was not included. All energy levels were calibrated by vacuum level and checked by the corresponding core levels.

For the comparison details of band edges, we take the follow procedure. First, all systems of twisted bilayer were fully relaxed. Second, we calculated the all symmetry stacking orders and some asymmetric orders at twist angle $\theta = 0°$ with the interlayer distance of these systems fixed to the result of first step. Third, we selected the highest and the lowest energy of band edges among all stacking orders and the difference of them is the maximum VBM (or CBM) difference.

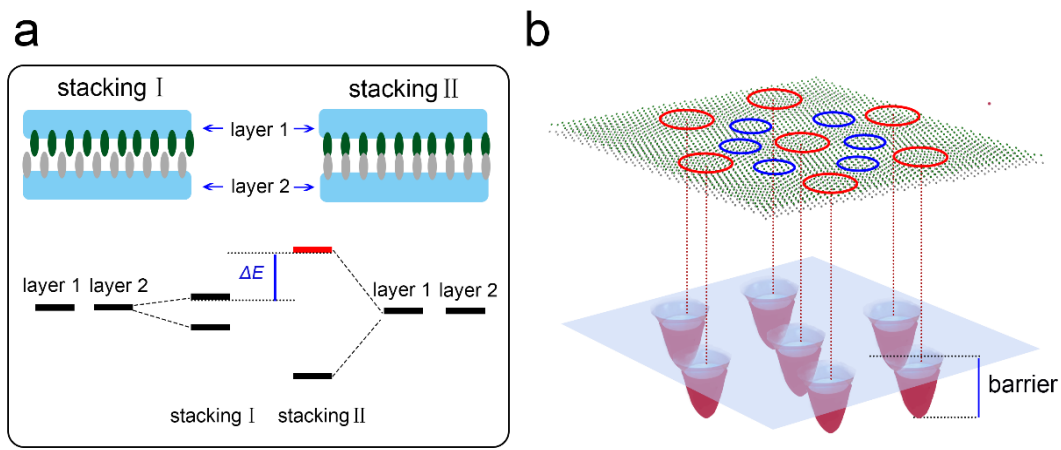

**Figure 1.** A scenario of the formation of energy battier. (a) Energy-level diagrams of the VBM for monolayer and bilayer with stacking I and stacking II neglecting the energy dispersion. And the VBM difference of stacking I and stacking II is indicated by $\varDelta E$. (b) The sketch map of twisted bilayer with its energy barrier. The blue and red circles highlight the stacking I and stacking II regions. And green and gray balls are from layer 1 and layer 2, respectively.

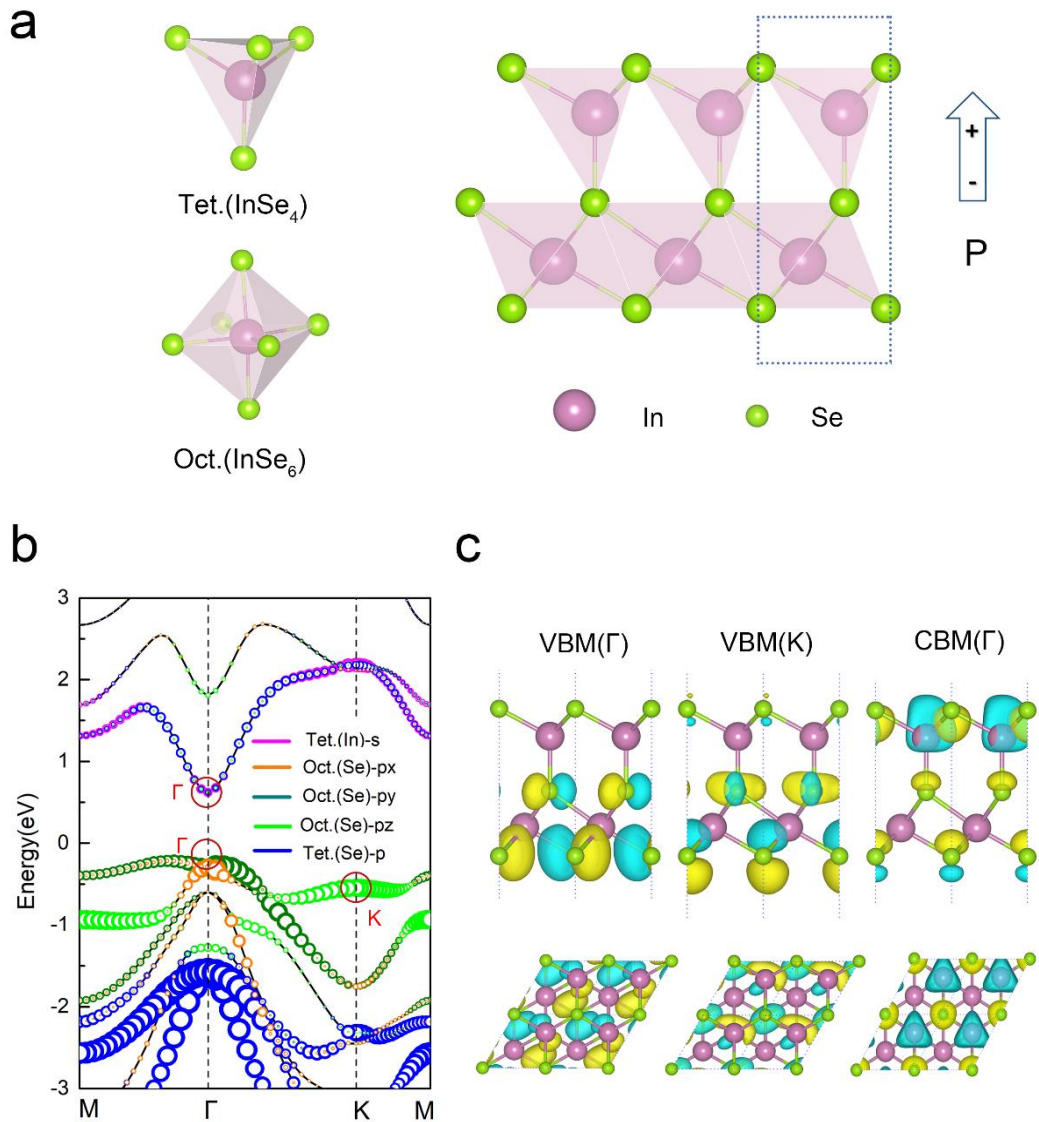

**Figure 2.** The geometric and band structure of the monolayer α-In$_2$Se$_3$. (a) The geometric structure of the monolayer α-In$_2$Se$_3$, which is constituted of InSe$_4$ tetrahedron and InSe$_6$ octahedral. And the unit-cell is framed in the dotted lines. The arrow indicates the direction of polarization. (b) The band structure of monolayer α-In$_2$Se$_3$. The contribution of In-$s$ and Se-$p$ orbitals at tetrahedron side is indicated by magenta and blue circles, respectively. The contribution of Se-$p_x$, Se-$p_y$ and Se-$p_z$ orbitals at octahedral side is indicated by orange, olive and green circles, respectively. (c) The side and top view of spatial distribution of wave function of the valence band edge at $k$ = Γ, $k$ = K and the conduction band edge at $k = \Gamma$ in Brillouin zone, as marked by the red circles in (b). Blue and yellow colors represent the phase of wave function. The isosurface value is $1.6 \times 10^{-6}$ e/bohr$^3$.

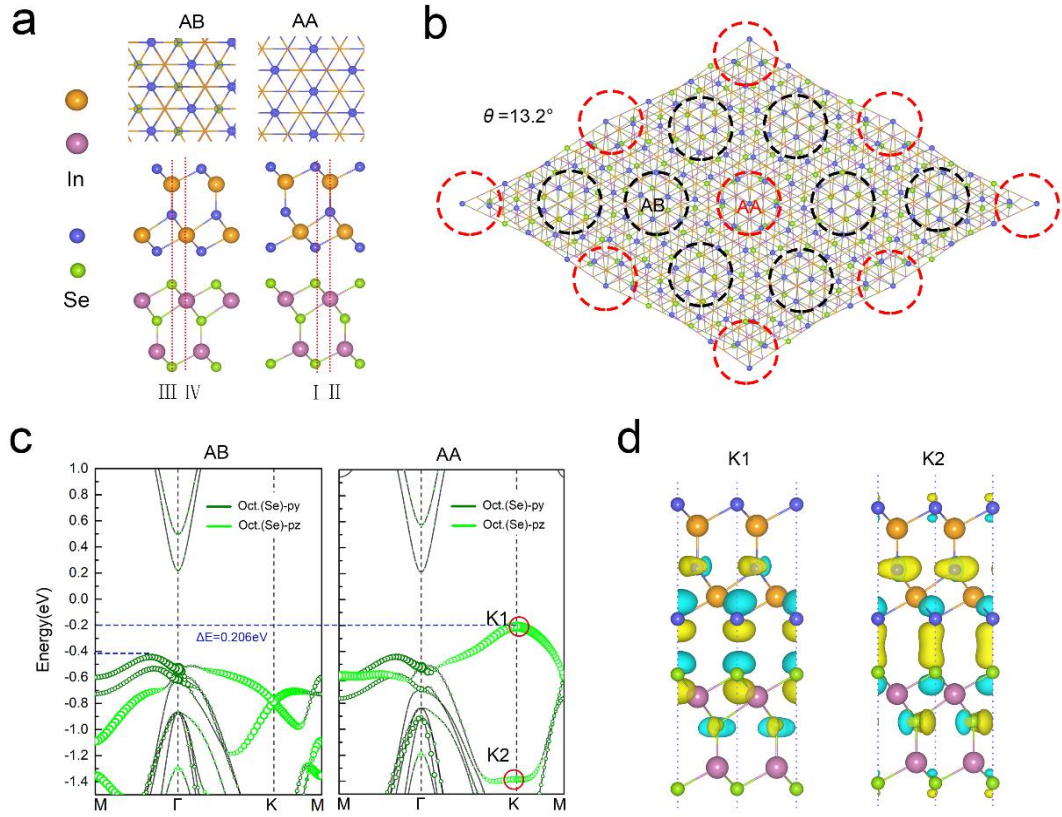

**Figure 3.** The geometric and band structure of the bilayer α-In$_2$Se$_3$. (a) Bilayer α-In$_2$Se$_3$ with AA stacking and AB stacking having octahedra face to each other. The dash red lines indicate the rotation axes of twisted bilayer α-In$_2$Se$_3$, as marked by I, II, III and IV. (b) Twisted bilayer α-In$_2$Se$_3$ with a twist angle $\theta$ =13.2°. The red and black circles highlight the AA stacking and AB stacking regions. (c) Energy bands of bilayer α-In$_2$Se$_3$ with AB stacking and AA stacking. The dash blue lines indicate the height of VBM, and the VBM difference of AA stacking and AB stacking is $\Delta E$=0.206eV. (d) The spatial distribution of wave function of the valence bands at $k$ = K1, $k$ = K2, as marked by the red circles in **c**. The isosurface value is $1.6 \times 10^{-6}$ e/bohr$^3$.

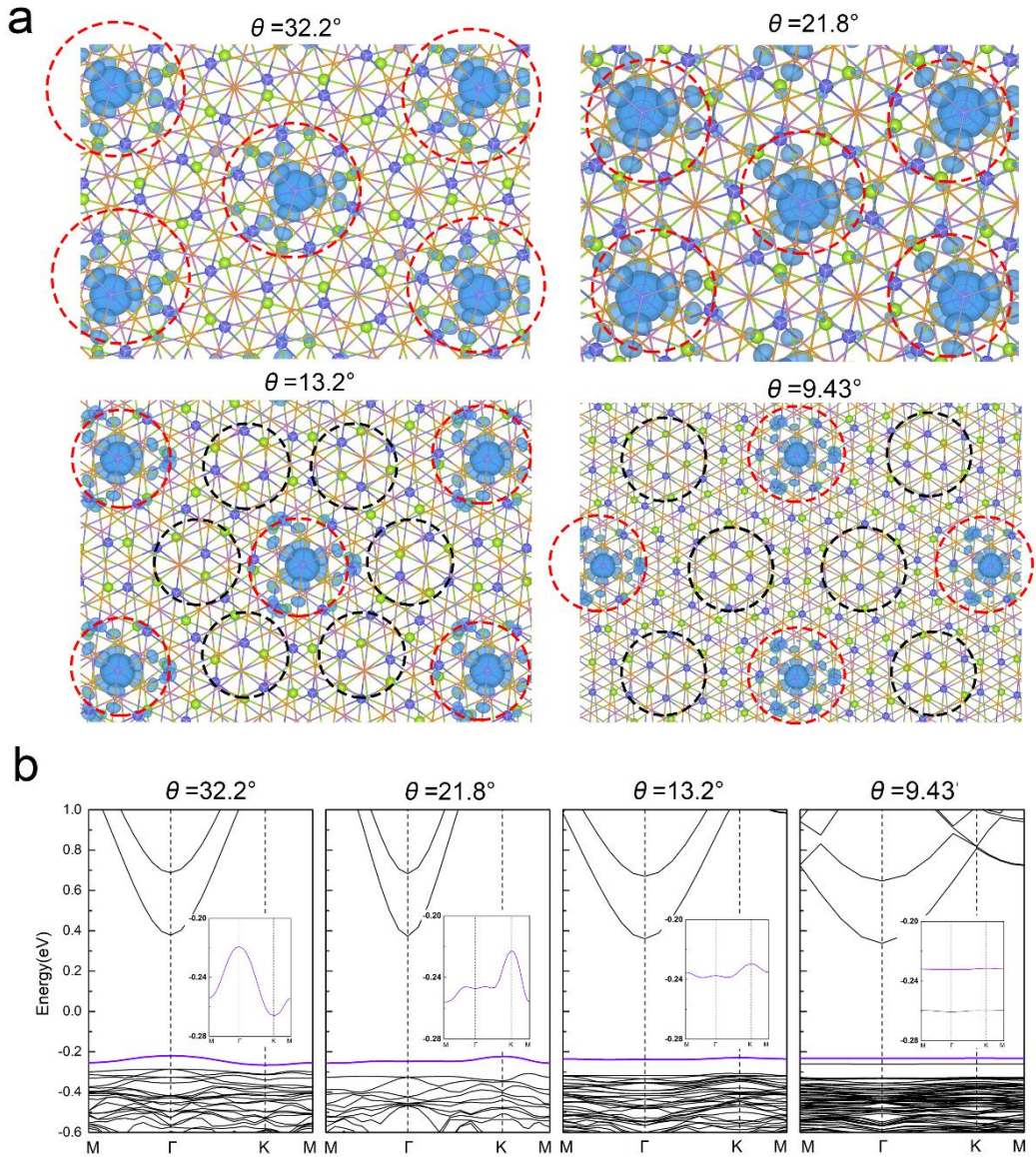

**Figure 4.** Flat bands at different twisted angles. (a) Twisted bilayer α-In$_2$Se$_3$ having octahedra face to each other at different twisted angles $\theta$ with the spatial distribution of wave function of the valence band edge at $k = \Gamma$. The red and black circles highlight the AA stacking and AB stacking regions. The isosurface value is 0.001 e/bohr$^3$. (b) The corresponding energy bands of twisted α-In$_2$Se$_3$ with different twisted angles $\theta$. The VBM highlighted by violet lines are magnified in the insets.

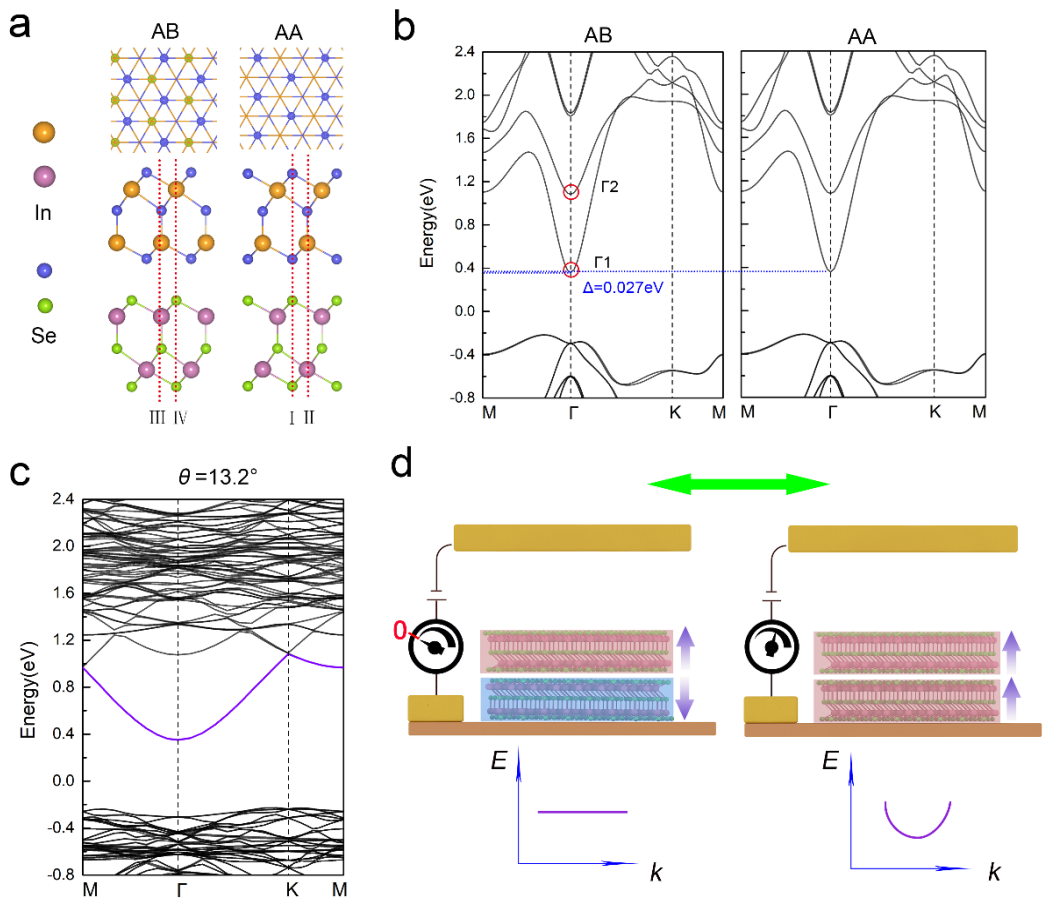

**Figure 5.** Control the flat bands by *E* field. (a) Bilayer α-In$_2$Se$_3$ with AA stacking and AB stacking having the tetrahedra face to each other. The dash red lines indicate the rotation axes of twisted bilayer α-In$_2$Se$_3$. (b) Energy bands of bilayer α-In$_2$Se$_3$ with AB stacking and AA stacking. The dash blue lines indicate the level of CBM, and the CBM difference of AA stacking and AB stacking is *ΔE*=0.027eV. (c) Bands of twisted bilayer α-In$_2$Se$_3$ with a twist angle *θ* =13.2° and (d) The sketch map of a dual gated device of twisted bilayer α-In$_2$Se$_3$. The regions of pink and blue bulks indicate the polarization up and down.

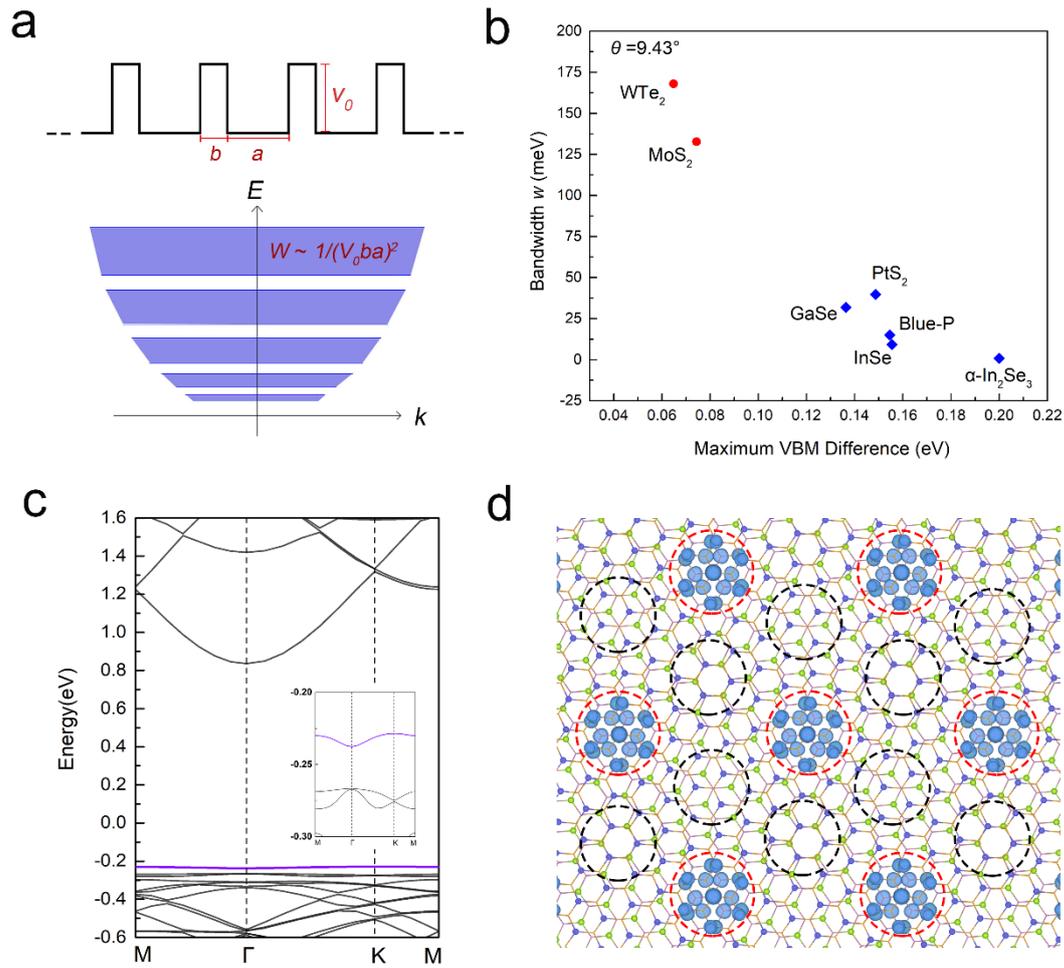

**Figure 6.** A new way to search the flat-band characters in moiré structures. (a) Kronig-Penny model diagrams. It is found that the bandwidth $W \sim 1/(V_0 ba)^2$, where $V_0$ is the height of barriers, $b$ and $a$ are width of barriers and wells respectively. (b) The bandwidth $W$ of the topmost valence band of a serials of bilayer vdW semiconductor with twist angle $\theta = 9.43°$ as a function of maximum VBM difference of different stacking orders. The systems are marked next to dots (here, α-$In_2Se_3$ refers to the octahedra in each layer facing to each other). (c) Energy bands of bilayer InSe with twist angle $\theta = 9.43°$. The VBM highlighted by violet lines are magnified in the insets. (d) The top view of spatial distribution of wavefunction of the valence band edge at $k = \Gamma$ in (c). The red and black circles highlight the AA stacking and AB stacking regions. The isosurface value is 0.0005 e/bohr$^3$.

## ASSOCIATED CONTENT

**Supporting Information**

The Supporting Information is available free of charge at

## AUTHOR INFORMATION

**Corresponding Author**

**Yunhao Lu** − *Zhejiang Province Key Laboratory of Quantum Technology and Device, Department of Physics, Zhejiang University, Hangzhou 310027, China*

**Authors**

**Shengdan Tao** − *Zhejiang Province Key Laboratory of Quantum Technology and Device, Department of Physics, Zhejiang University, Hangzhou 310027, China*

**Xuanlin Zhang** − *State Key Laboratory of Silicon Materials, School of Materials Science and Engineering, Zhejiang University, Hangzhou 310027, China*

**Jiaojiao Zhu** − *Research Laboratory for Quantum Materials, Singapore University of Technology and Design, Singapore 487372, Singapore*

**Pimo He** − *Zhejiang Province Key Laboratory of Quantum Technology and Device, Department of Physics, Zhejiang University, Hangzhou 310027, China*

**Shengyuan A. Yang** − *Research Laboratory for Quantum Materials, Singapore University of Technology and Design, Singapore 487372, Singapore*

**Su-Huai Wei** − *Beijing Computational Science Research Center, Beijing 100193, China*

Complete contact information is available at

**Author Contributions**

Y.L. proposed, designed the project and organized the paper. S.T., X.Z. and J.Z. conducted the simulations. Y.L., S.Y. made the data analyses and wrote the manuscript. S.W. and P.H. contributed to the discussion of results and manuscript revision.

**Notes**

The authors declare no competing financial interest.

## ACKNOWLEDGMENTS

This work was financially supported by the National Key R&D Program of China (2019YFE0112000), Zhejiang Provincial Natural Science Foundation of China (LR21A040001), National Natural Science Foundation of China (11974307, 12088101, 11991060, U1930402), and Singapore MOE AcRF Tier 2 (MOE2019-T2-1-001).